\documentclass[graybox]{svmult}

\usepackage{mathptmx}       
\usepackage{helvet}         
\usepackage{courier}        
\usepackage{type1cm}        
\usepackage{makeidx}         
\usepackage{graphicx}        
\usepackage{multicol}        
\usepackage[bottom]{footmisc}

\newcommand{\betatwo}{$\frac{\beta^{2}}{2}$}

\makeindex             

\begin{document}

\title*{
Underground test of quantum mechanics - the VIP2 experiment
}
\author{Johann Marton, S. Bartalucci, A. Bassi, M. Bazzi, S. Bertolucci, C. Berucci, M. Bragadireanu, M. Cargnelli, A. Clozza, C. Curceanu, L. De Paolis, S. Di Matteo, S. Donadi, J.-P. Egger, C. Guaraldo, M. Iliescu, M. Laubenstein, E. Milotti, A. Pichler, D. Pietreanu, K. Piscicchia, A. Scordo, H. Shi, D. Sirghi F. Sirghi, L. Sperandio, O. Vazquez-Doce, E. Widmann and J. Zmeskal}
\institute{Johann Marton \at Stefan Meyer Institute, Boltzmanngasse 3, 1090 Vienna, Austria \\ \email{johann.marton@oeaw.ac.at}
\and S. Bartalucci \at LNF-INFN, Via Enrico Fermi 40, Frascati, Italy
\and A. Bassi \at  Dipartimento di Fisica, Universita di Trieste and INFN– Sezione di Trieste, Via Valerio, 2, I-34127 Trieste, Italy
\and M. Bazzi \at  LNF-INFN, Via Enrico Fermi 40, Frascati, Italy
\and S. Bertolucci \at CERN, Route de Meyrin 385, 1217 Meyrin, Switzerland
\and C. Berucci \at Stefan Meyer Institute, Boltzmanngasse 3, 1090 Vienna, Austria
\and M. Bragadireanu \at “Horia Hulubei” National Institute of Physics and Nuclear Engineering, Str. Atomistilor no. 407, P.O. Box MG-6, Bucharest - Magurele, Romania
\and M. Cargnelli \at Stefan Meyer Institute, Boltzmanngasse 3, 1090 Vienna, Austria
\and A. Clozza \at LNF-INFN, Via Enrico Fermi 40, Frascati, Italy
\and C. Curceanu \at LNF-INFN, Via Enrico Fermi 40, Frascati, Italy 
         \at Museo Storico della Fisica e Centro Studi e Ricerche “Enrico Fermi”, Roma, Italy \\ \email{catalina.curceanu@lnf.infn.it}
\and L. De Paolis \at LNF-INFN, Via Enrico Fermi 40, Frascati, Italy
\and S. Di Matteo \at Institut de Physique UMR CNRS-UR1 6251, Universit´e de Rennes1, F-35042 Rennes, France
\and S. Donadi \at Dipartimento di Fisica, Universita di Trieste and INFN– Sezione di Trieste, Via Valerio, 2, I-34127 Trieste, Italy
\and J.-P. Egger \at Institut de Physique, Universit´e de Neuchˆatel, 1 rue A.-L. Breguet, CH-2000 Neuchˆatel, Switzerland
\and C. Guaraldo \at LNF-INFN, Via Enrico Fermi 40, Frascati, Italy
\and M. Iliescu \at LNF-INFN, Via Enrico Fermi 40, Frascati, Italy
\and M. Laubenstein \at  INFN, Laboratori Nazionali del Gran Sasso, S.S. 17/bis, I-67010 Assergi (AQ), Italy
\and E. Milotti \at Dipartimento di Fisica, Universita di Trieste and INFN– Sezione di Trieste, Via Valerio, 2, I-34127 Trieste, Italy
\and A. Pichler \at Stefan Meyer Institute, Boltzmanngasse 3, 1090 Vienna, Austria \\ \email{andreas.pichler@oeaw.ac.at}
\and D. Pietreanu \at LNF-INFN, Via Enrico Fermi 40, Frascati, Italy
\and K. Piscicchia \at LNF-INFN, Via Enrico Fermi 40, Frascati, Italy
        \at Museo Storico della Fisica e Centro Studi e Ricerche “Enrico Fermi”, Roma, Italy 
\and A. Scordo \at LNF-INFN, Via Enrico Fermi 40, Frascati, Italy
\and H. Shi \at LNF-INFN, Via Enrico Fermi 40, Frascati, Italy
\and D. Sirghi \at LNF-INFN, Via Enrico Fermi 40, Frascati, Italy
\and F. Sirghi \at LNF-INFN, Via Enrico Fermi 40, Frascati, Italy
\and L. Sperandio \at LNF-INFN, Via Enrico Fermi 40, Frascati, Italy
\and O. Vazquez-Doce \at Excellence Cluster Universe, Technische Universit\"at M\"unchen, Garching, Germany
\and E. Widmann \at Stefan Meyer Institute, Boltzmanngasse 3, 1090 Vienna, Austria
\and J. Zmeskal \at Stefan Meyer Institute, Boltzmanngasse 3, 1090 Vienna, Austria
}

\titlerunning{The VIP2 experiment}
\authorrunning{Johann Marton et al.}
\maketitle

\abstract{
We are experimentally investigating possible violations of standard quantum mechanics predictions in the Gran Sasso underground laboratory in Italy. We test with high precision the Pauli Exclusion Principle and the collapse of the wave function  (collapse models).
We present our method of searching for possible small violations of the Pauli Exclusion Principle (PEP) for electrons, through the search for {\it anomalous} X-ray transitions in copper atoms, produced by {\it fresh} electrons (brought inside the copper bar by circulating current) which can have the probability to undergo Pauli-forbidden transition to the 1 s level already occupied by two electrons and we describe the VIP2 (VIolation of PEP) experiment under data taking at the Gran Sasso underground laboratories.
In this paper the new VIP2 setup installed in the Gran Sasso underground laboratory will be presented. The goal of VIP2 is to test the PEP for electrons with unprecedented accuracy, down to a limit in the probability that PEP is violated at the level of 10$^{-31}$. We show preliminary experimental results and discuss implications of a possible violation.}


\section{Introduction}
The Pauli Exclusion Principle (PEP) is a fundamental principle in physics, valid for identical-fermion systems. It forms the basis of: the periodic table of elements, electric conductivity in metals and the degeneracy pressure which makes white dwarf stars and neutron stars stable. Furthermore it is a consequence of the Spin-Statistics connection \cite{Pauli1940} and is embedded into the quantum field theory \cite{Luders1958}.\\
Despite the fact that the PEP is connected to so many fundamental phenomena, an intuitive explanation is still missing \cite{Feynman1963}. Moreover, in the framework of theories beyond the Standard Model, a violation of the PEP might occur (e.g. \cite{Jackson2008a}). Nowadays, the interest in quantum foundations increased dramatically \cite{Khrennikov2012, d2014preface}. Furthermore, recent work on Spin-Statistics has been carried out in \cite{Santamato2015, DeMartini2014}. Thus, it is important to test the PEP for each fermionic particle type. In the last two decades, many experiments have been carried out, which set upper limits for the probability of its violation ( \cite{Bernabei2010}, \cite{Bellini2010}, \cite{Hilborn1996}, \cite{Barabash2010}, \cite{Tsipenyuk1998}, \cite{Nolte1991} and \cite{Abgrall2016}). These results were primarily obtained as by-products of experiments with a different main scientific objective (like BOREXINO \cite{Bellini2010} and DAMA \cite{Bernabei2010}). As some of these experiments are investigating the validity of the PEP for composite particles like nucleons and nuclei, it is important to note that the VIP2 experiment investigates atomic transitions of electrons, which are elementary particles.\\
%
The different approaches to investigate the PEP need to be distinguished concerning their possible fulfillment of the Messiah-Greenberg (MG) superselection rule (\cite{Messiah1964}, \cite{Elliott2012}). This rule states that the symmetry of the wavefunction of a steady state is constant in time. As a consequence, the symmetry of a quantum state can only change if a particle, which is new to the system, interacts with the state. All of the aforementioned experiments are looking for changes in the symmetry of steady states that would be violating the MG superselection rule.\\
\section{Tests of the Pauli Exclusion Principle}
\begin{figure}[h]
 \centering
 \includegraphics[width=0.9\textwidth]{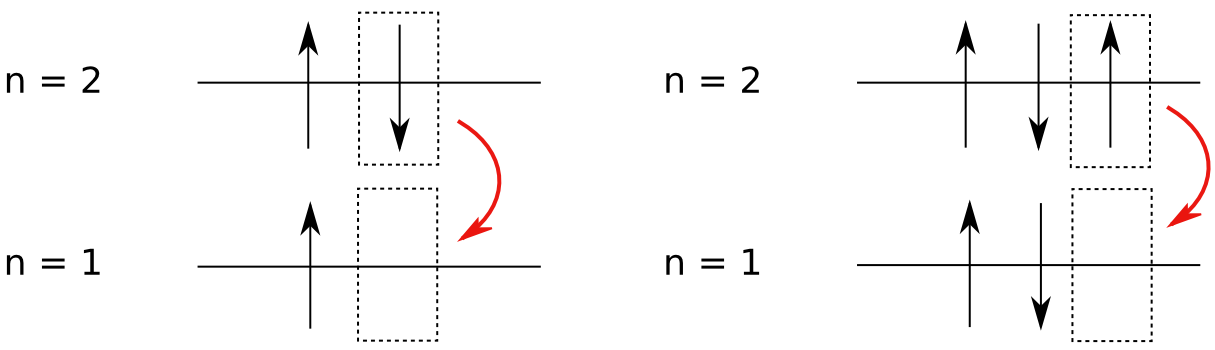}
 \caption{Normal atomic 2p to 1s transition with an energy of 8.05 keV in copper (left) and the corresponding non-Paulian transition with an energy of around 7.7 keV in copper (right).}
 \label{fig:energy-scheme}
\end{figure}

\begin{figure}[h]
 \centering
 \includegraphics[width=0.8\textwidth]{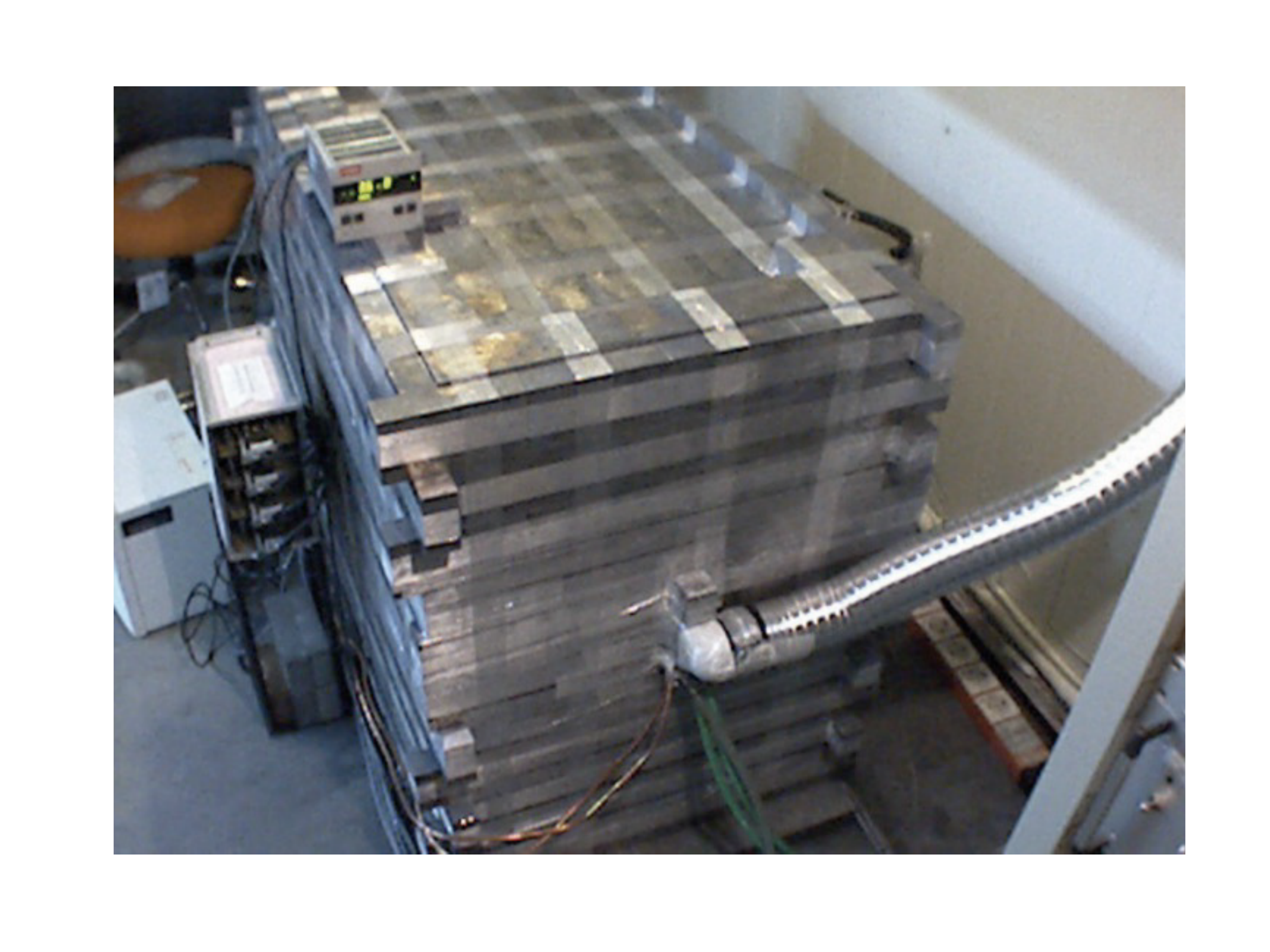}
 \caption{The VIP experiment with passive shielding mounted taking data at LNGS.}
 \label{fig:lead-coffin}
\end{figure}

One of the first experiments looking for a small violation of the PEP was conducted by Goldhaber and Scharff-Goldhaber in 1948 \cite{Goldhaber1948}. It was originally designed to check if the particles that made up beta rays were the same as the electrons in atoms, but it was later used to put an upper bound to the probability of the violation of the Pauli exclusion principle. In this experiment, beta rays were absorbed by a block of lead. The idea of the authors was, that if the 2 kinds of particles were not identical, the beta ray particle could be captured by the atom and cascade down to the ground state without being subject to the PEP. The X-rays emitted during this cascation process were recorded and used to set upper bounds for a violation of the PEP.\\
To the best of our knowledge, the best way to circumvent the MG superselection rule and test the PEP with high precision is to introduce ``new'' electrons in a conductor via a current. The electrons form new quantum states with the atoms in the conductor. The goal is to search for new quantum states, which have a symmetric component in an otherwise antisymmetric state. These non-Paulian states can be identified by the characteristic radiation they emit during atomic transitions to the ground state.\\
The first to employ this scheme in a pioneering experiment in 1988 were E. Ramberg and G. A. Snow \cite{Ramberg1990}. The experiment searched for X-rays originating from Pauli-forbidden atomic transitions, in this case from the 2p to the fully occupied 1s ground state. These transitions are depicted in figure \ref{fig:energy-scheme}.
The ``new'' electrons introduced by the current can be seen as test particles, as they can be used to study interactions between a fermionic system and a fermion which has not previously interacted with the studied system. The experiment of Ramberg and Snow set an upper limit for the probability that the PEP is violated for electrons of $\frac{\beta^{2}}{2} <$ 1.7 x $10^{-26}$. The parameter \betatwo is quasi standard in the literature for the probability that the PEP is violated.\\
A much improved version of the experiment of Ramberg and Snow was set up by the VIP collaboration \cite{Collaboration2004}. It employed Charge-Coupled Devices (CCDs) as soft X-ray detectors and, through careful selection of the involved materials and shielding, a reduction of background was achieved. The VIP experiment, conducted at the underground laboratory Laboratori Nazionali del Gran Sasso (LNGS) in Italy, took data for $\sim$3 years until 2010. The sensitivity of the experiment greatly increased due to the reduction of background induced by cosmic rays. This background is reduced by 6 orders of magnitude at LNGS compared to experiments above ground. The experiment set a preliminary upper limit for the probability that the PEP is violated for electrons of $\frac{\beta^{2}}{2} <$ 4.7 x $10^{-29}$ (\cite{Curceanu2011a}, \cite{Pietreanu2014}). A picture of the experiment can be seen in figure \ref{fig:lead-coffin}.\\
A similar experiment of this type conducted in recent years by colleagues in the US, with a prototype for the MAJORANA demonstrator, is described in \cite{Elliott2012}. It covers the same topic as the VIP experiment, but uses a complementary apparatus. This common interest with the VIP collaboration in testing fundamental physics shows the interest of the scientific community in foundations of quantum mechanics and theories beyond the Standard Model of particle physics. The most recent experiment in this field is VIP2, which is the subject of this proposal. It was in detail described in recent publications, for example \cite{Pichler2016, Shi2016, Marton2015, Curceanu2016a}. It is the follow-up experiment of VIP. For VIP2, several crucial components were upgraded, like the target, X-ray detectors and shielding.


\section{VIP 2 at LNGS underground laboratory}

\begin{figure}[h]
 \centering
 \includegraphics[width=0.85\textwidth]{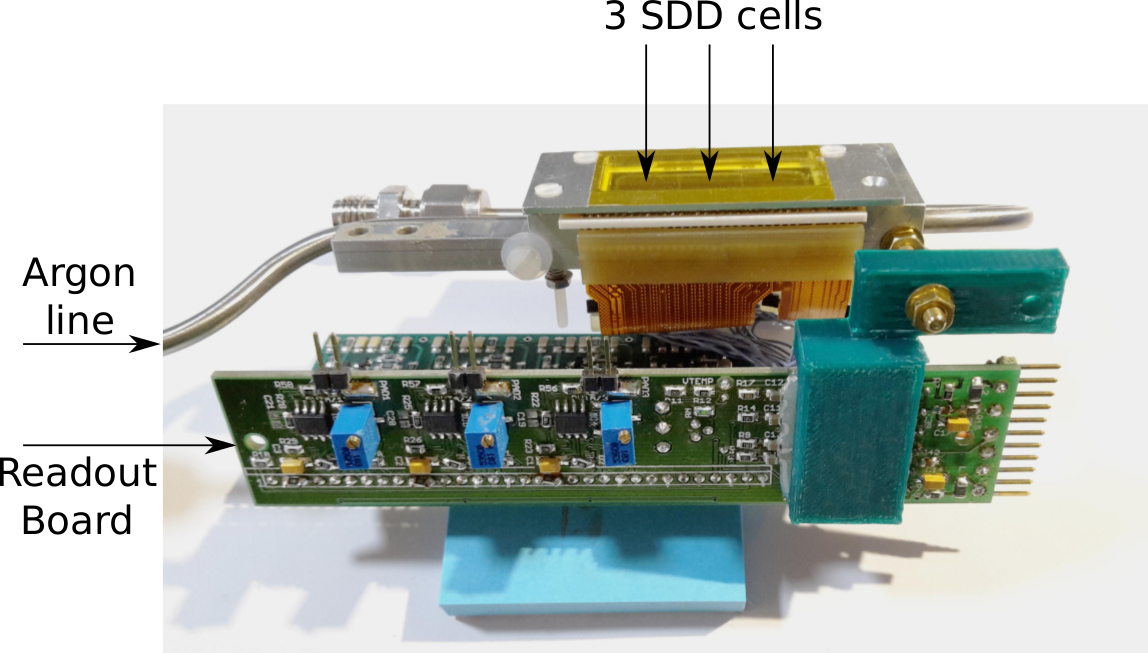}
 \caption{Silicon Drift Detectors with cooling and readout electronics.}
 \label{fig:SDDs}
\end{figure}

\begin{figure}[h]
 \centering
 \includegraphics[width=0.8\textwidth]{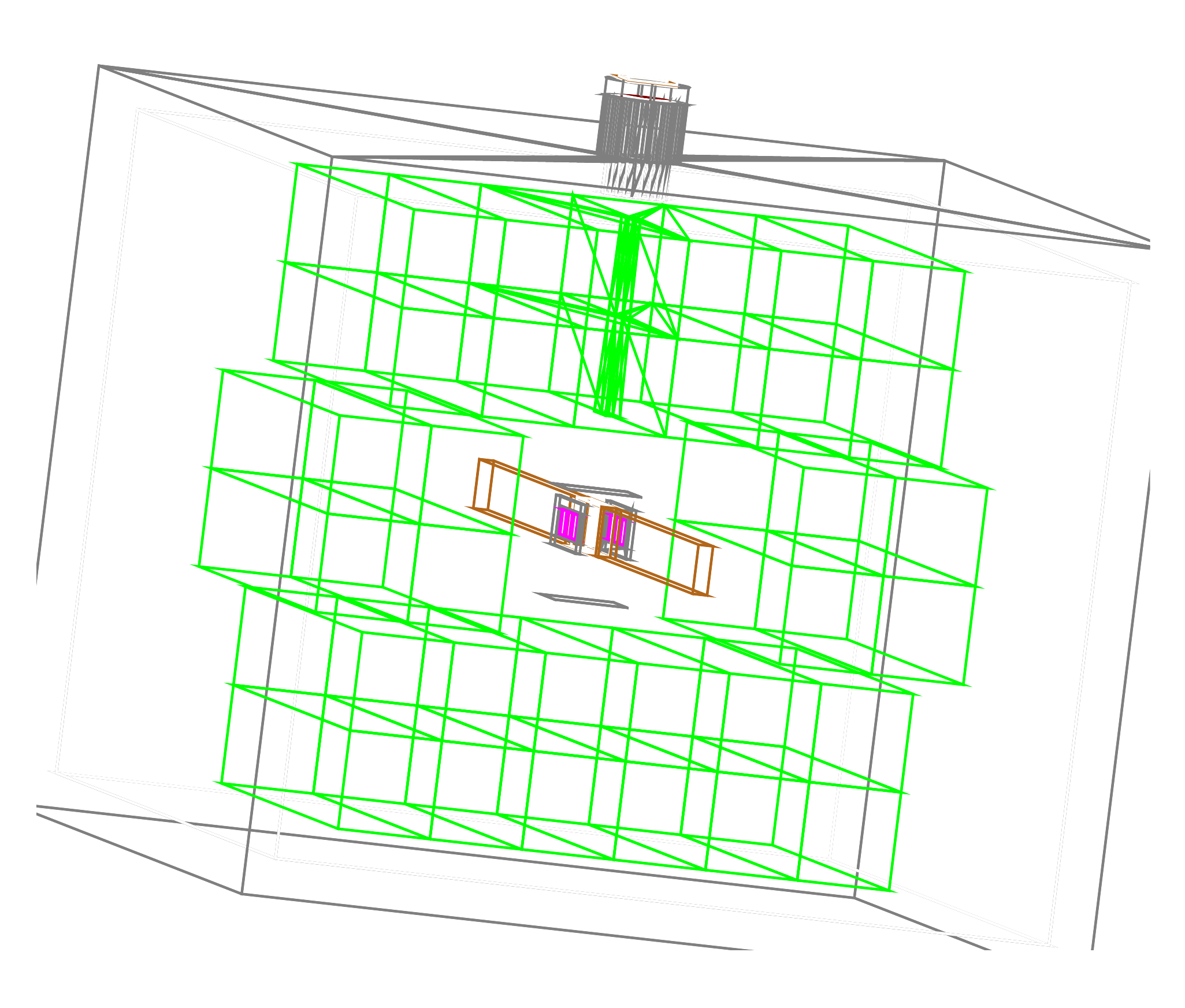}
 \caption{The setup modeled in GEANT4, with the scintillators (green), the copper conductor (brown) and the SDDs (purple). The aluminum enclosure is also shown (grey).}
 \label{fig:MC-setup}
\end{figure}

\begin{figure}[h]
 \centering
 \includegraphics[width=0.9\textwidth]{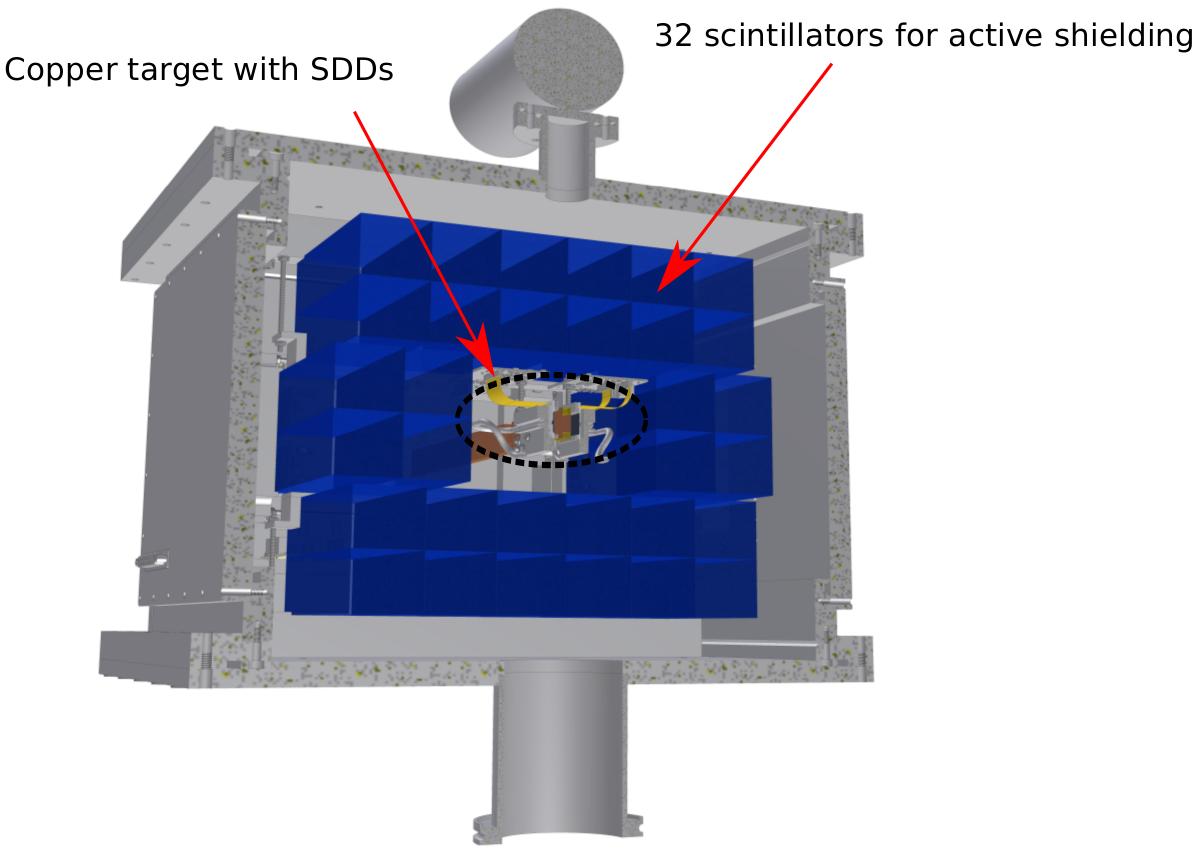}
 \caption{A render of the VIP2 setup including the Silicon Drift Detectors and the active shielding system.}
 \label{fig:active-shielding}
\end{figure}

\begin{figure}[h]
 \centering
 \includegraphics[width=\textwidth]{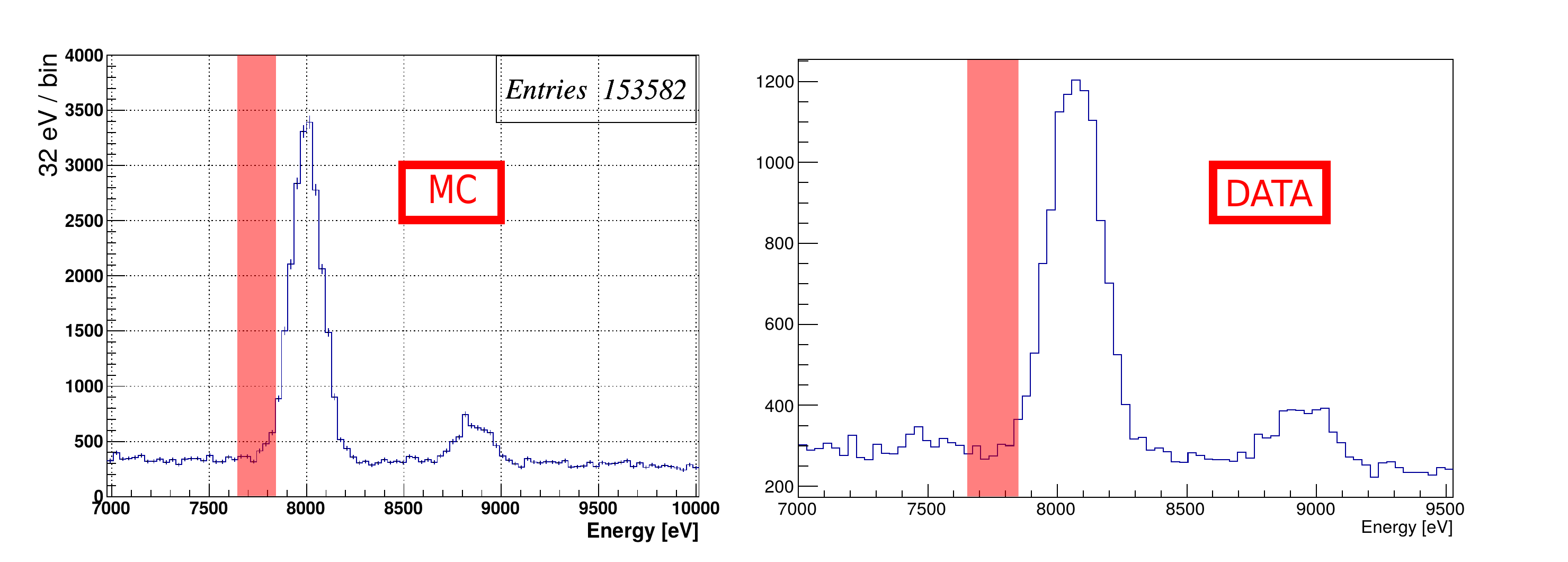}
 \caption{Comparison between 30 days of Monte Carlo simulation data (left) and 30 days of data measured at LNGS (right). The background in the region of interest (marked in red) differs only by around 30 \%.}
 \label{fig:mc-data}
\end{figure}

The VIP2 experiment is taking data at LNGS in Italy. Conducting the experiment at this facility is advantageous, because of its low-background environment. The Gran Sasso laboratory is the facility of this kind which is easiest to reach for the experimenters from Stefan Meyer Institute, as there is no laboratory of this kind in Austria. In 2016, we took data for 4 months at LNGS.\\ \\
The core part of the setup are the SDDs which are used as soft X-ray detectors (\cite{Cargnelli2005}, \cite{Lechner}). The experiment utilizes 6 SDD cells with an active area of 1 cm$^{2}$ each. The cells are located on each side of the ultra pure copper target, where the high current runs through. The target consists of two copper strips with a gap of 6 mm between the strip and the respective SDD array. Each of these strips has a length of 91 mm and a width of 20 mm. With this configuration, the SDDs cover a solid angle of $\sim$7 $\%$ of the target. The probability for detecting an X-ray originating from the target is then further reduced from this value by X-ray attenuation in the copper strip. The heating of the target due to the high current is counteracted by water cooling. The water line runs between the two strips and keeps the copper strip below room temperature, even with a current of 100 A. The SDDs are cooled by liquid argon to a temperature of 100 K. The whole experimental setup is evacuated to approximately $10^{-5}$ mbar, in order to enable the SDD cooling at 100 K. A picture of the SDDs with the liquid argon cooling line and readout electronics is shown in figure \ref{fig:SDDs}.
The increase in signal strength (i.e. the amount of detected X-rays from non-Paulian transitions per time) gained by upgrading the VIP experiment to the setup described above is summarized in table \ref{tab:VIP2gain}.
\begin{table}[h!]
 \centering
\begin{tabular}{|c|c|c|c|} 
\hline
  & VIP & VIP2 & Gain factor\\
\hline
 geometry & 0.021 \cite{Bartalucci2006} & 0.03 & 3/2\\
\hline
 detector efficiency & 0.48 & 0.99 & 2 \\
\hline
 current & 40 A & 100 A & 5/2\\
\hline 
\hline
 \textbf{Total} & & & \textbf{7 - 8}\\
\hline
\end{tabular}
\caption{The gain factors for increasing the signal strength in the VIP2 experiment compared to the preceding VIP experiment are given in the table. They are in agreement with the original proposal for VIP2 \cite{Marton}.} 
\label{tab:VIP2gain}
\end{table}
The factor in the first line describes the probability that a Pauli-forbidden X-ray produced in the target passes through a SDD. It includes effects of target and SDD geometry as well as X-ray absorption in the target. This factor was increased by mounting the SDDs closer to the target than the CCDs of VIP, which increases the solid angle covered by the detectors. These figures are verified by GEANT4 \cite{Agostinelli2003} based Monte Carlo simulations (M. Cargnelli, private communication, 2016). For this purpose and all other mentioned GEANT4 simulations, the complete setup was modeled in this framework. A picture of the simulated setup is shown in figure \ref{fig:MC-setup}. 
\\
The second gain factor represents the higher X-ray detection efficiency of SDDs compared to CCDs. It comes from the fact that the depth of the depletion layer of CCDs is 30 $\mu$m \cite{Zmeskal}, whereas the depletion layer of SDDs is 450 $\mu$m thick. The difference in depths results in a difference in quantum efficiency of a factor of 2. The measurements for VIP2 can be undertaken with a higher current of 100 A due to the new copper target geometry and the implemented water cooling. \\
Overall, these factors increase the signal by around one order of magnitude. This enhancement factor is in agreement with the VIP2 proposal \cite{Marton}. All the mentioned parts have been tested successfully in the laboratory at the Stefan Meyer Institute in Vienna and at LNGS.\\
%
The energy and the time resolution of the SDDs are core properties of the experiment. The detector performance which was anticipated in \cite{Marton} has been verified experimentally. The energy resolution was determined to be around 150 eV (FWHM), tested with an Fe-55 source at 6 keV, for all 6 SDDs. The time resolution was measured to be around 400 ns (FWHM) relative to a scintillator trigger, which exceeds the original target \cite{Marton}.\\
As an active shielding system, we use an assembly of 32 plastic scintillators read out by Silicon Photomulitpliers (SiPMs). They are arranged around the copper target and the SDDs. The purpose of the active shielding system is to reject all SDD events which coincide with events in the scintillators, as these are caused by radiation originating from outside of the setup. Making this time coincidence is only possible due to the good time resolution of the SDDs. A render of the copper target with the active shielding system is shown in figure \ref{fig:active-shielding}.
The detection efficiency of the active shielding system was determined to be around 97 $\%$ for 500 MeV electrons at the beam test facility at the DA$\Phi$NE collider at the Laboratori Nazionali di Frascati (LNF) in Italy. Tests in the laboratory at the Stefan Meyer Insitute (SMI) in Vienna showed that the detection efficiency is around 95 $\%$ for the given cosmic ray background. \\
The cosmic ray background at LNGS is lower than at SMI by about 6 orders of magnitude. The main source of background at LNGS are high energy photons in the range of around 40 - 500 keV, for which the detection efficiency of the active shielding system is around 5 $\%$. This was predicted by recent Monte Carlo simulations which were based on a scintillator detection threshold of 100 keV deposited energy. This is the energy equivalent of the voltage threshold used in the experiment. Further reducing the threshold is not possible due to unavoidable noise in the detection system. The result from simulations was confirmed by data taken at LNGS in 2016. The simulations lead to a quantitative understanding of the background induced by the gamma radiation reported in \cite{Haffke2011}. A comparison between the simulated and the measured spectra are shown in figure \ref{fig:mc-data}.

\section{VIP2 Results and future plans}

In 2016, we were able to take a total of 40 days of data with a current of 100 A and 70 days of data without current. Using an analysis technique analogous to the one used by Ramberg and Snow \cite{Ramberg1990} on this data set, we are able to set a preliminary upper limit for the probability that the PEP is violated in the electron sector of \betatwo $\leq$ 1,4 x $10^{-29}$. This result represents the most stringent test of the PEP in a system circumventing the MG superselection rule.

\subsection{The planned upgrade}

\begin{figure}[h]
  \centering
  \includegraphics[width=0.75\textwidth]{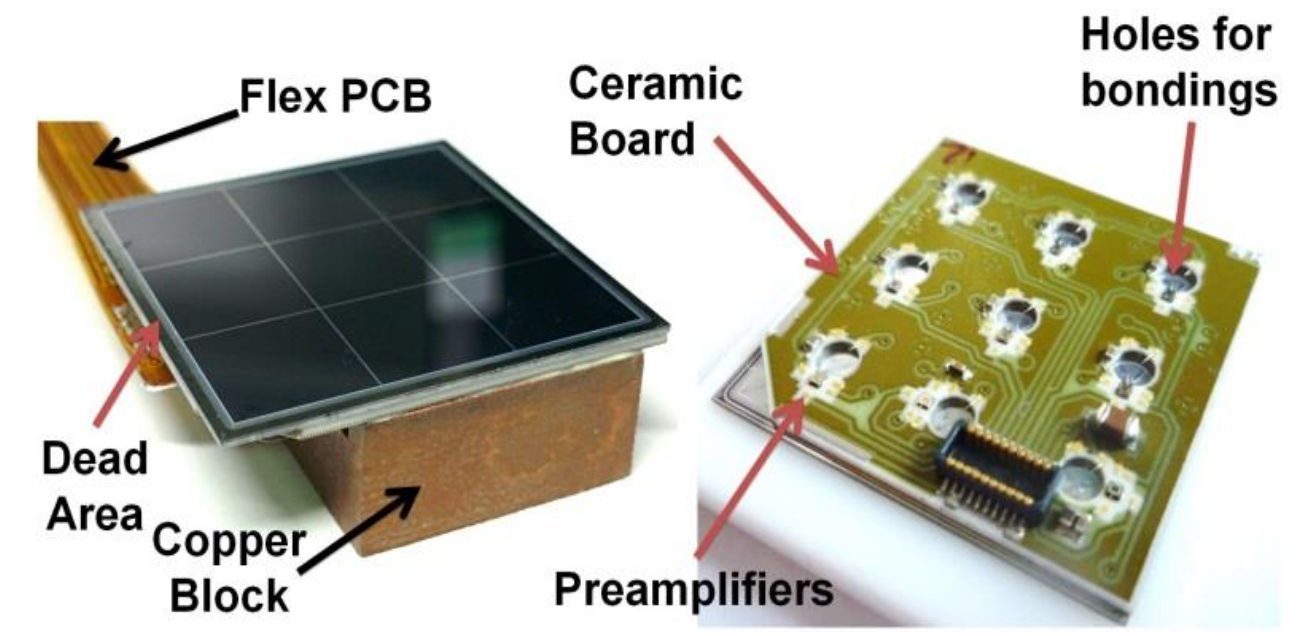}
  \caption{New type of SDD with ceramic board for contacting and readout and a copper block for cooling.}
  \label{fig:3x3SDD}
\end{figure}

\begin{figure}[h]
 \centering
 \includegraphics[width=0.8\textwidth]{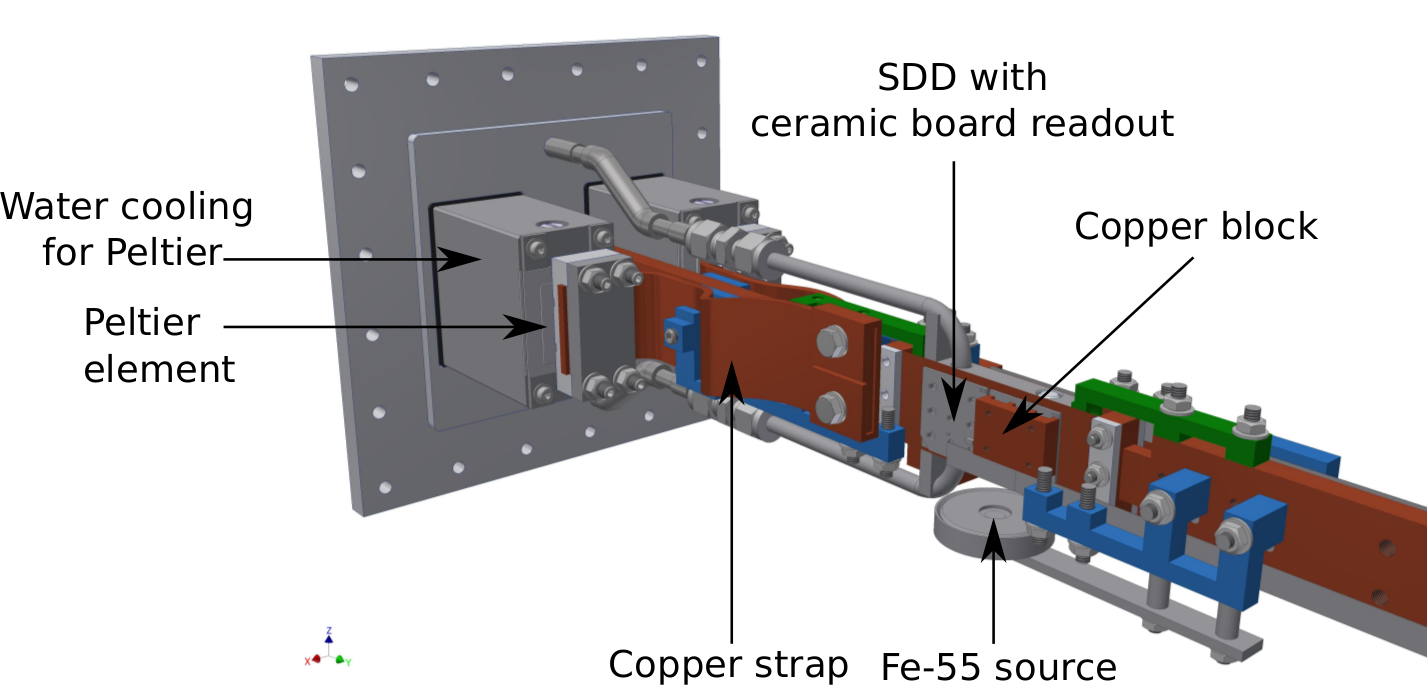}
 \caption{Render of the upgraded setup: The Peltier element is attached to a copper block, which is attached to the backside of the ceramic board, with a copper strap. The Fe-55 source for energy calibration is shown. Some parts of the setup are not displayed to enhance the visibility.}
 \label{fig:peltier-cooling-setup}
\end{figure}

We are planning to further enhance the signal and reduce the background in the energy region of the forbidden transition. Together, these effects will improve the upper limit on the violation of the PEP we will be able to set after the running time of the experiment, by more than one order of magnitude.\\ \\
To reduce the background, it is important to shield the detector from high energy photon radiation. This will be done by a passive shielding consisting of two parts. An outer part, 5 cm in thickness, made of low radioactivity lead and an inner part which is 5 cm in thickness, made of low radioactivity copper. Both parts will completely enclose the setup. The inner copper part rests on a frame constructed from Bosch profiles. The frame and the brick layout are already planned. The geometry of the enclosure was optimized to reach maximum background suppression. The copper and lead blocks are available at LNGS and only need to be assembled. Due to our understanding of the origin of the background, and GEANT4 simulation results, we are confident that the installation of shielding will reduce the background in the energy region of interest by at least a factor 20. To further increase the passive shielding from the outside photon radiation in the energy region of the non-Paulian X-ray transition at 8 keV, a plan to include a Teflon shielding of approximately 5 mm thickness inside of the experimental setup around the copper target and the silicon detectors has been developed.\\
Another fundamental part of the optimized experiment will be the implementation of new SDDs \cite{Fiorini2013}. The new detectors were developed in a cooperation between SMI, Politecnico di Milano and the Fondazione Bruno Kessler (FBK). They consist of units of 9 single cells of 8 x 8 mm$^{2}$, assembled in a 3 x 3 matrix with a fraction of active area as high as 85 $\%$. A picture of the SDD unit is shown in figure \ref{fig:3x3SDD}.%
Four of the SDD units will be used, with two on each side of the target. With one cell having a surface area of 64 mm$^{2}$, the total active area will be around 23 cm$^{2}$, i.e. about four times the current active area of 6 cm$^{2}$. According to GEANT4 simulations, this leads to a higher detection rate of X-rays from non-Paulian transitions by a factor of 3. This is due to the increase in the solid angle coverage of the target. Another advantage is that this type of detector can be operated at higher temperatures of around 230 K. The currently used SDDs are operated at 100 K and require argon cooling. The higher operating temperature can be provided by Peltier cooling. Peltier cooling is better suited for long term data taking, because of its stable and failure-free operation. The setup for Peltier cooling and signal readout of the SDDs is displayed in figures \ref{fig:3x3SDD}, \ref{fig:peltier-cooling-setup}.\\
A ceramic board for the SDD voltage supply and the readout is mounted on the side of the SDDs opposite to the radiation entrance windows. The first stage of preamplification is provided by a new preamplifier (CUBE), which was recently developed by Politecnico di Milano. These preamplifiers allow high performance X-ray spectroscopy with standard SDD technology. The ceramic board is connected to a readout board for further amplification and data acquisition.\\
On the backside of the ceramic board a copper block is mounted which is attached to the cold side of a Peltier element. This attachment will be realized in the upgraded setup by a thermally conductive copper strap (see figure \ref{fig:peltier-cooling-setup}). It is via this copper strap, that the SDD is cooled by the Peltier element. The warm side of the Peltier element is cooled by a closed water cycle with a cooling pump. A similar water cooling system is currently in use to cool the copper target. This system can be adapted to cool the Peltier elements in addition to the copper target. This system of SDD combined with Peltier cooling has already been tested at the laboratory of the Stefan Meyer Insitute in Vienna. A typical energy resolution was found to be 200 eV (FWHM) at 6 keV.
\\
In order to reduce the background coming from radioactive radon, the whole setup, including the passive shielding, will be enclosed in an existing plastic box where nitrogen is flushed. This Radon Reduction System (RRS) reduces the radon concentration in the atmosphere surrounding the experiment. Radon is an important source of background at LNGS, as it is part of the decay chains of uranium and thorium, which in turn are abundant in the rocks of the Gran Sasso mountains. \\
\subsection{Gain for the VIP2 experiment}
\begin{figure}[h]
 \centering
 \includegraphics[width=0.8\textwidth]{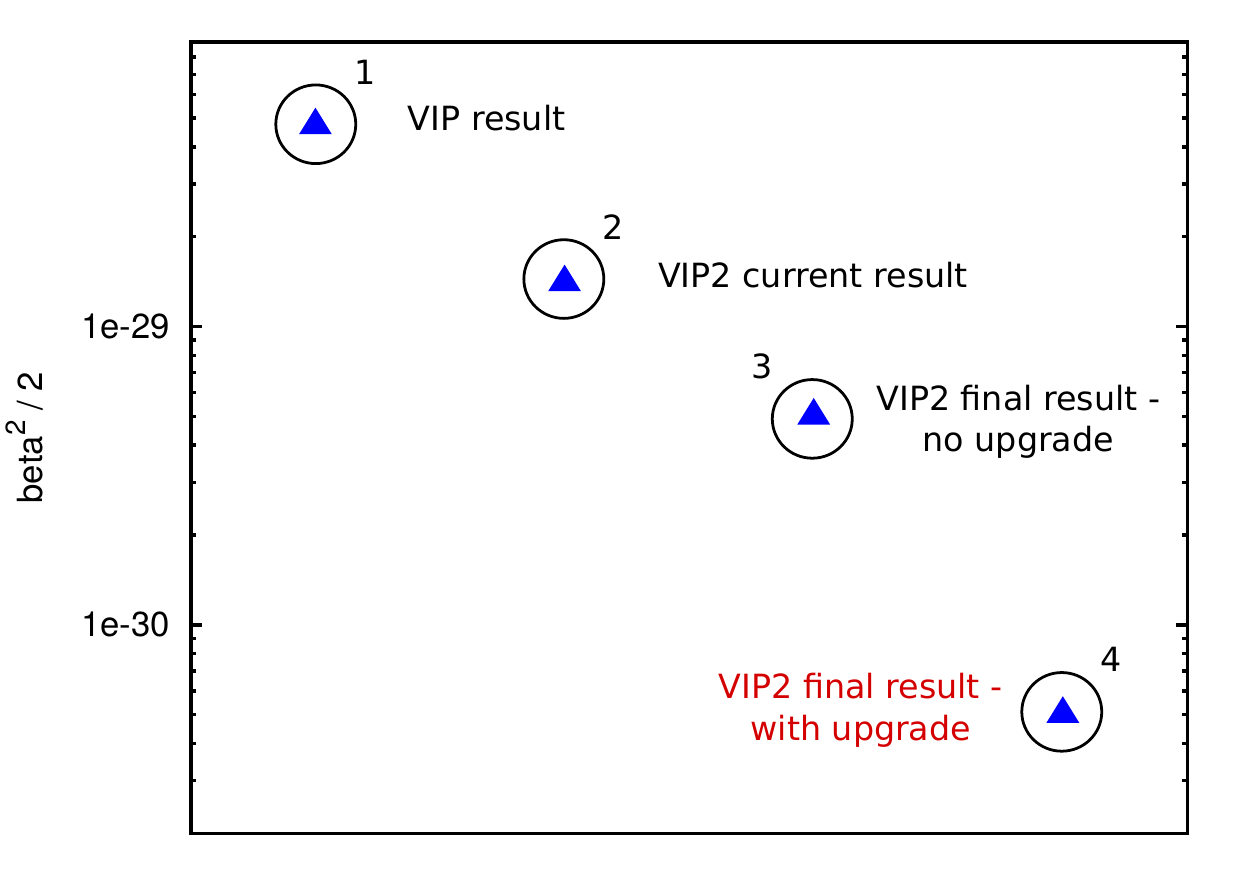}
 \caption{The preliminary upper limit for the violation of the Pauli Exclusion Principle obtained by data taken with VIP2 (2) compared to the preliminary result of the former VIP experiment (1) and different scenarios (3) and (4) (see text).}
 \label{fig:beta-value}
\end{figure}

The mentioned upgrade will improve the final achievable value for $\frac{\beta^{2}}{2}$ by at least one order of magnitude compared to the final value achievable with the current setup. The upgrades are summed up in table \ref{tab:VIP2-upgrade}. 
\begin{table}[h!]
 \centering
\begin{tabular}{|c|c|c|c|} 
\hline
 Upgrade & Signal enhancement & Background reduction & Gain\\
\hline
 new SDDs & 3 &  $\sim$ 0.45 & $\sim$ 4/3\\
\hline
 Passive Shielding & - & $\geq\sqrt{20}$ & $\geq$ 4.5 \\
\hline
 RRS & - & $\sqrt{3}$ & $\sqrt{3}$\\
\hline
\hline
 \textbf{Total Gain} & & & \textbf{$\geq$ 10}\\
\hline
\end{tabular}
\caption{Factors contributing to the improvement of the sensitivity of the VIP2 experiment (see text).} 
\label{tab:VIP2-upgrade}
\end{table}
\\
In the first line the effects of the new SDDs are listed. They will enhance the signal (i.e. the number of possible detected X-rays from non-Paulian transitions per time) by a factor of at least 3, due to their larger solid angle coverage. Due to their larger area, they will also increase the background counts by a factor of $\frac{23}{6}$. Additionally, the anticipated energy resolution of around 200 eV (FWHM) will enlarge the background in the region of interest by around a factor of $\frac{4}{3}$. Since the background enters as a square root into the calculations of \betatwo, this brings the total gain for the upper limit for a PEP violation to approximately $\frac{4}{3}$. An additional advantage of the new detectors which can not be put in this table is the easier handling, as mentioned earlier. The Peltier cooling replaces all the parts needed for cooling with closed cycle liquid argon cooling, e.g. a helium compressor with coldhead and condenser which are used to liquefy the argon, an electronic argon temperature controller and the argon cooling line inside the setup. The Peltier cooling is advantageous due to its easier handling and long term stability.\\
The lead and copper shielding outside of the setup, and Teflon around the detectors inside of the setup, will reduce the background by at least a factor of 20. This corresponds to a gain for \betatwo of around 4.5, which has been verified with GEANT4 simulations.\\
The nitrogen flushed around the setup to decrease the radon concentration (RRS) will reduce the background by a factor of around 3. As a result, the gain in sensitivity will be about $\sqrt{3}$. Together this adds up to an improvement of at least one order of magnitude, shown in figure \ref{fig:beta-value}.
In the figure, the points represent the following from left to right: (1) the preliminary value obtained with the complete data set of the predecessor experiment VIP; (2) the preliminary value  from the data taken until end of 2016 with VIP2; (3) the expected VIP2 value after around 3 more years of data taking with the current setup. Finally (4) corresponds to the expected value which can be achieved after three years running time with the planned upgrades. \\ \\
\section{Conclusion and Outlook}
We will be able to install the upgrade by the end of 2017 after thorough tests of the detectors at SMI. Thereby the VIP2 experiment will be able to set a new upper limit for the probability that the PEP is violated in the order of $10^{-31}$ by the end of the running time of the experiment. Compared to the preliminary result of the VIP experiment of \betatwo $\leq$ 4.7 x $10^{-29}$, this is an improvement by more than 2 orders of magnitude. The new value will also improve the current value set by VIP2 by more than one order of magnitude and will represent a test of the PEP with unprecedented sensitivity.
\\ \\ \\
\textbf{Acknowledgements}\\ \\
We thank H. Schneider, L. Stohwasser, and D. Pristauz-Telsnigg from Stefan-Meyer-Institut for their
fundamental contribution in designing and building the VIP2 setup.
We acknowledge the very important assistance of the INFN-LNGS laboratory staff during all
phases of preparation, installation and data taking.
The support from the EU COST Action CA 15220 is gratefully acknowledged.
We thank the Austrian Science Foundation (FWF) which supports the VIP2 project with the
grants P25529-N20 and  W1252-N27 (doctoral college particles and interactions) and Centro Fermi for the grant “Problemi aperti nella meccania quantistica”.
Furthermore, this paper was made possible through the support of a grant from the
Foundational Questions Institute, FOXi (“Events” as we see them: experimental test of the
collapse models as a solution of the measurement problem) and a grant from the John Templeton
Foundation (ID 581589). The opinions expressed in this publication are those of the authors
and do not necessarily reflect the views of the John Templeton Foundation.\\
%

%
%

\end{document}